\documentclass[12pt, a4paper, oneside]{article}

\usepackage{amsmath}
\usepackage{amssymb}
\usepackage{latexsym,bm}

\usepackage{graphicx}
\usepackage{wrapfig}
\usepackage{multirow}
\usepackage{subfig}

\usepackage[inner=2cm,outer=2cm, top=3cm, bottom=3.2cm]{geometry}

\usepackage{natbib} 
\bibpunct{(}{)}{;}{a}{,}{,}
\usepackage[colorlinks=true,linkcolor=black,citecolor=blue]{hyperref}

\begin{document}

%%%% Make title  %%%%%%%%%%%%%%%%%%%%%%%%%%%%%%%%%
\title{\textbf{Inertial wave and zonal flow in librating spherical shells}}
\author{Yufeng Lin$^1$, Jerome Noir$^1$, Michael A. Calkins$^2$ \\
        \small{1. Institute of Geophysics, ETH Zurich} \\
        \small{2. Department of Earth and Space Sciences, UCLA}}
\date{\today}
\maketitle

%%%% Abstract %%%%%%%%%%%%%%%%%%%%
\begin{abstract}
We numerically study the inertial waves and zonal flows in spherical shells driven by longitudinal libration, an oscillatory variation of rotation rate. Internal shear layers are generated due to breakdown of the Ekman boundary layer at critical latitudes. Our numerical results validate the scaling laws of internal shear layers predicted by previous studies. Mean zonal flows are driven by the non-linear interaction in the boundary layers. Non-linear interaction of inertial waves in the interior fluids has no significant contribution to the zonal flow. Multiple geostrophic shear layers are generated due to non-linearities in the boundary layers at critical latitudes and reflection points of internal shear layers. We also investigate the scaling laws of geostrophic shear layers and extrapolate the results to the planetary setting.   

\end{abstract}
%%%%%%%%%%% Main Body

\section{Introduction}
Many planetary bodies in our solar system possess large volume of fluids either in the form of subsurface ocean or liquid iron core \citep{Anderson1996,Comstock2003,Margot2007}. Perturbation flows in the planetary fluids can be driven by many mechanisms. Longitudinal libration, an oscillatory variation of the rotation rate, is one of them. Many planetary bodies undergo forced longitudinal libration \citep{Comstock2003}, hereafter referred to as libration. Understanding the fundamental dynamics and physics of these fluid systems has significant geophysical or astrophysical implications as they might result in energy dissipation and magnetic field generation of planetary bodies. In this paper will numerically study the inertial waves and zonal flows in librating spherical shells.

In rotating fluid systems fluid motion in the form of oscillations are referred to as inertial waves \citep{Greenspan1968}. Inertial waves can be sustained in different ways. Laboratory experiments \citep{Aldridge1969,Noir2009} and numerical simulations \citep{Tilgner1999,Calkins2010} have showed that libration can excite axisymmetrical inertial waves. However, inertial waves can survive only for the libration frequency less than twice of the background rotation rate. The corresponding inviscid linear equations are then hyperbolic with conical characteristic surfaces. The angle between characteristic surfaces and rotation axis, defined as $\alpha$, depends on the forcing frequency $f$ and is given by \citep{Greenspan1968}
\begin{equation}
f=2\sin\alpha.
\end{equation}
%The inertial waves propagate along the characteristic surfaces.

A viscous boundary layer, usually refer to as Ekman boundary layer, is formed in order to satisfy no-slip boundary condition on solid boundary. The Ekman boundary layer has thickness of $E^{1/2}$ and the mass flux between Ekman boundary layers and interior fluids also scales as $E^{1/2}$ normally. Here $E$ is Ekman number, which is the ratio of typical viscous force to the Coriolis force. However, these scalings break down at a certain critical latitude where the energy carried by incident inertial waves is reflected tangent to boundary instead of back into interior fluids. In spherical geometry the critical latitude is determined as $\theta_c=\arcsin(f/2)$ \citep{Greenspan1968}.
The thickness of Ekman boundary layer increases to $E^{2/5}$ at critical region over a width of $E^{1/5}$ \citep{Roberts1963}. Internal shear layers, the superposition of inertial waves, are spawned from critical latitudes propagating along the characteristic surface in the interior fluids due to the Ekman boundary breakdown \citep{Kerswell1995a}. \citet{Calkins2010} summarized the amplitude and structure of the internal shear and the boundary layers in a rotating spherical shell (see their Fig. 1). In this paper, we will numerically validate the scaling laws of internal shear layers proposed by previous studies. 

Mean zonal flows can be generated due to the non-linear interaction in rotating fluids \citep{Busse1968,Tilgner2007,Noir2010}. In librating systems there are three possible mechanism to drive the zonal flows: (i) non-linear interaction of sidewall boundary instabilities, (ii) non-linear self-interaction of inertial waves in the interior and (iii) non-linearities in the Ekman boundary layer \citep{Noir2010}. Both numerical \citep{Calkins2010} and experimental \citep{Noir2010} results showed that the Taylor-Gortler vortices due to sidewall instabilities do not drive zonal flow. \citet{Greenspan1969} proved that non-linear interaction of purely inviscid inertial modes does not produce significant zonal flow and conjectured the non-linear effect within the viscous boundary layer is the essential mechanism to drive the steady zonal flow. However, tidal driven zonal flow studies showed that viscosity modified inertial modes can drive significant zonal flows \citep{Tilgner2007, Morize2010}. In the limit of low libration frequencies, \citet{Busse2010} derive an analytical solution of zonal flow driven by non-linearity in viscous boundary layer in low frequency librating spherical cavity. The solution predicts that the zonal flow scales with the square of the libration amplitude. This has been confirmed by the laboratory and numerical experiments \citep{Sauret2010}. In precession study, \citet{Busse1968} showed that the non-linearities in the critical Ekman boundary layer can produce a differential rotation of steady flow. We refer to the differential rotation of steady flow as the geostrophic shear layer. Our numerical results will show that the zonal flows are primarily driven by non-linearities in the Ekman boundary layer. Non-linear interaction of inertial waves in the interior has no significant contribution on zonal flows. Not only the breakdown of the Ekman boundary at critical latitude can produce a geostrophic shear layer, but the reflection of internal shear layers on the Ekman boundary layer also can. We will demonstrate how the different geostrophic shear layers scale as Ekman number.   

\section{Method}
We consider a rapidly rotating spherical shell of outer radius $r_o$ and inner radius $r_i$, enclosed with homogeneous, Newtonian fluid. The radius ratio $\eta=r_i/r_o$ is the only non-dimensional geometry parameter to define a spherical shell. In the reference frame rotating at background rate $\Omega_o$, the inner or outer boundary is librating as
\begin{equation}
\omega(t)=\Delta \phi \omega_L \sin(\omega_Lt),
\end{equation}
where $\Delta\phi$ is libration amplitude and $\omega_L$ is angular libration frequency. Using $\Omega_o^{-1}$ and $r_o$ as the typical time scale and length scale, the non-dimensionalized governing equation in the rotating frame can be written as \citep{Noir2010}
\begin{equation}
\frac{\partial \mathbf u}{\partial t}+\mathbf{u}\cdot \nabla\mathbf{u}+2\mathbf{\hat z}\times\mathbf u=
-\nabla p+E\nabla^2\mathbf u,
\end{equation}
\begin{equation}
\nabla \cdot \mathbf u=0,
\end{equation}
where $\mathbf u$ is velocity; $\mathbf{\hat z}$ is unit vector parallel to the rotation axis; $p$ is modified pressure including the centrifugal term; $E=\dfrac{\nu}{\Omega_or_o^2}$ is Ekman number which measures the ratio of viscous force and Coriolis force.

The non-dimensionalized no-slip boundary condition for the inner core librating is
\begin{equation}
u_\phi(\eta,\theta,\phi)=\eta \varepsilon \cos \theta \sin(ft),
\end{equation} 
and for the outer shell librating is 
\begin{equation}
u_\phi(1,\theta,\phi)=\varepsilon \cos \theta \sin(ft).
\end{equation} 

The non-dimensional libration amplitude, $\varepsilon=\Delta \phi f$, measures the relative strength of the librational forcing to the background solid rotation, and $f=\dfrac{\omega_L}{\Omega_o}$ is the ratio of angular frequency of libration and the background rotation. 

The fully non-linear equations are solved numerically using a spectral code developed by \citet{Calkins2010}. 

\section{Results}
\subsection{Inertial wave}
For small libration amplitude the time dependent flows in a librating spherical shell are dominated by inertial waves \citep{Noir2010}. Our numerical results show that internal shear layers, superposition of inertial waves, are generated due to the breakdown of the critical latitude and propagate along the characteristic surfaces. We will validate the scaling laws of the internal shear layer predicted by previous studies, which are basement for non-linear interaction and mean zonal flows analysis. 
\subsubsection{Internal shear layers spawned from inner critical latitude}
\begin{figure}
\begin{center}
\subfloat[]{\includegraphics[scale=0.5]{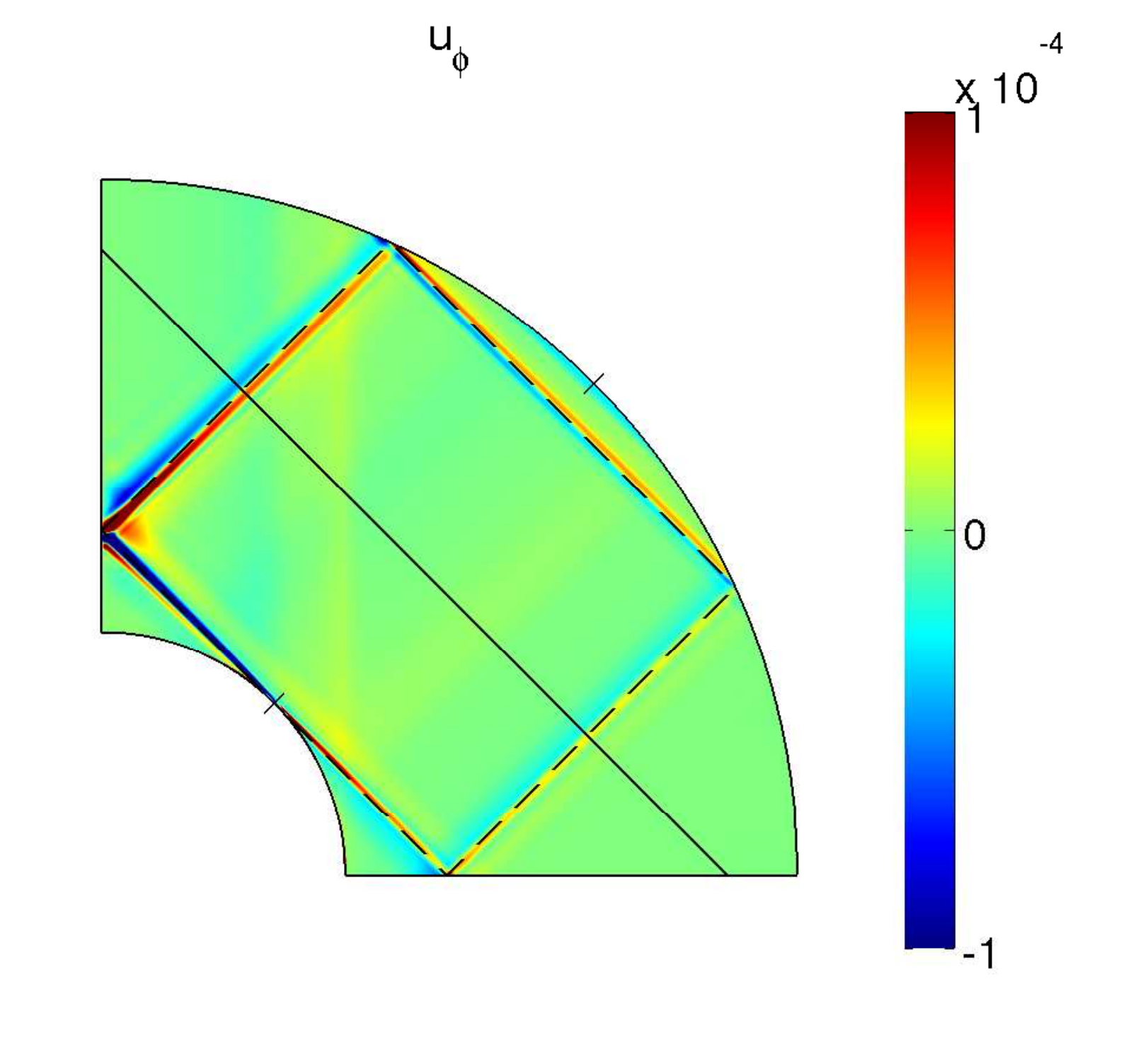}}
\hspace{0.2cm}
\subfloat[]{\includegraphics[scale=0.45]{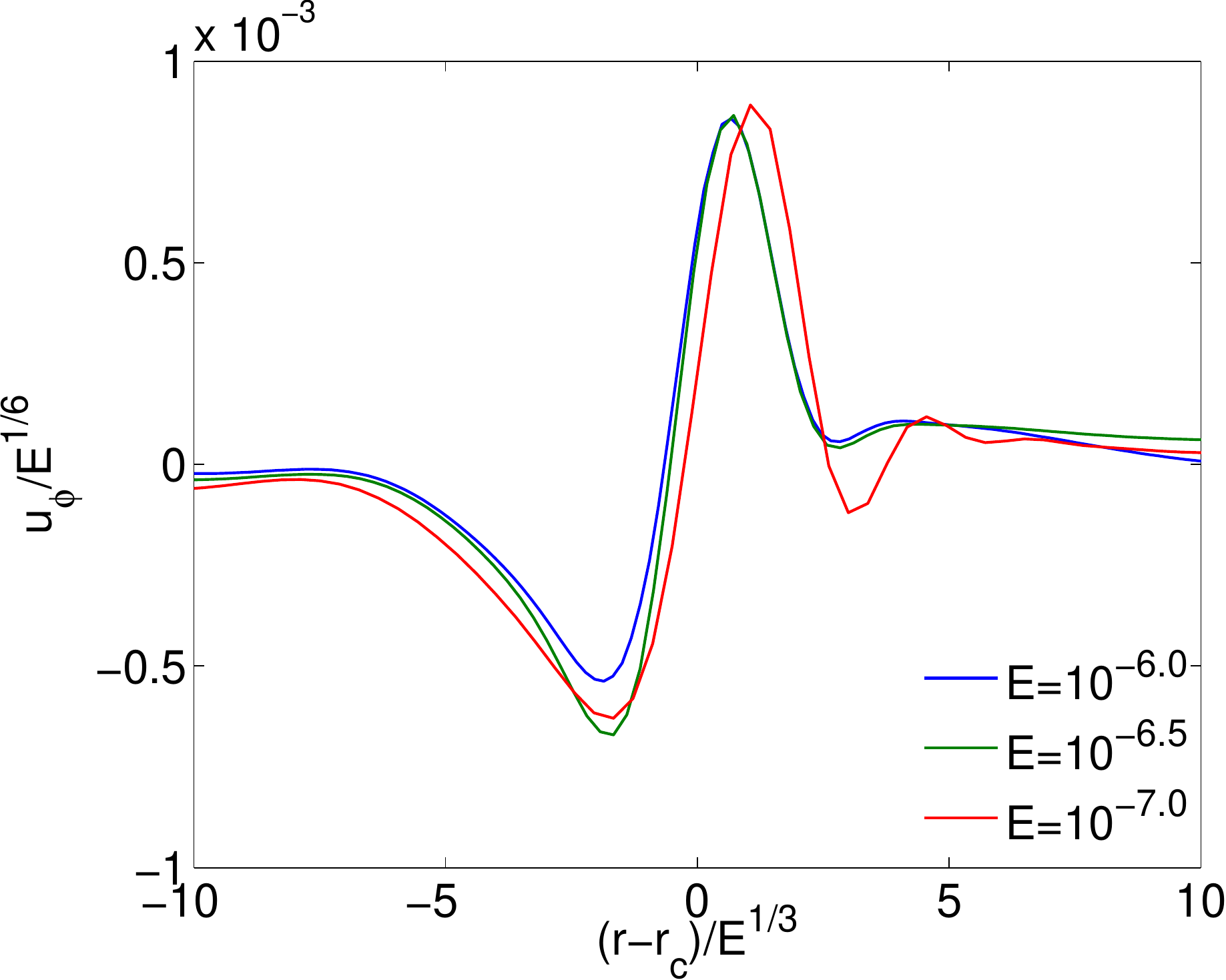}}
\caption{Azimuthal velocities for librating inner core and uniformly rotating outer shell. (a) Azimuthal velocities in a meridional plane when the inner shell is at the maximum retrograde position. Tick marks show the location of the critical latitude and dashed lines show the ray path of inertial wave exited from critical latitude. $E=10^{-7.0}, f=1.4142, \varepsilon=1.4142\times10^{-2}, \eta=0.35$. (b) Velocity profiles along a line (solid line in (a)) across the internal shear layer for different Ekman numbers. The coordinate origin is set at the center of the shear layer, the distance and velocity is normalized by $E^{1/3}$ and $E^{1/6}$ respectively.}
\label{inner_shear}
\end{center}
\end{figure}

First we consider the inner core is librating at frequency $f=1.4142$ with amplitude $\varepsilon=1.4142\times 10^{-2}$ and the outer shell is uniformly rotating at $\Omega_o$. The radius ratio $\eta=0.35$.  \autoref{inner_shear} (a) shows azimuthal velocities in a meridional plane when the inner core is at the maximum retrograde position for Ekman number $10^{-7}$. We can see that internal shear layers are spawned from inner critical latitude propagating along the characteristic surfaces (dashed lines in \autoref{inner_shear} (a)). When an internal shear layer meets a boundary, it is reflected back into interior. 
%As we mentioned previously, the angle between characteristic surface and the rotation axis is only determined by libration frequency.
Because the libration frequency is not modified by a reflection, internal shear layers are reflected such that the angle enclosed with the rotation axis remains unchanged. Due to this peculiar reflection law, it is possible to form closed path or so-called attractor after multiple reflections \citep{Tilgner1999}. For $f=1.4142$, the angle between the characteristic surface and rotation axis is $45^\circ$. The internal shear layers spawned from inner critical latitude form a closed path after multiple reflections. 
%It should be mentioned that the "reflections" occurred on the rotation axis and equator attribute to the axisymmetry and equatorial symmetry, they are not physical reflection. 

%Thanks to simple path of the internal shear layers this special libration frequency, we can study the detailed  structure of the internal shear layers.
Azimuthal velocities are plotted along a line (solid line in \autoref{inner_shear} (a)) across the internal shear layer for Ekman numbers ranging from $10^{-6}$ to $10^{-7}$ in \autoref{inner_shear} (b). The coordinate origin is set at the center of the shear layer and the distance is normalized by $E^{1/3}$, meanwhile, the velocity amplitude is normalized by $E^{1/6}$. After normalization, these shear layers for different Ekman numbers are almost collapsed. Our numerical results indicate the amplitude and width of internal shear layer spawned from inner critical latitude scale as $E^{1/6}$ and $E^{1/3}$ respectively, which is consistent with previous theoretical prediction by \citet{Kerswell1995a}.

\subsubsection{Internal shear layers spawned from outer critical latitude}
\begin{figure}
\begin{center}
\subfloat[]{\includegraphics[scale=0.5]{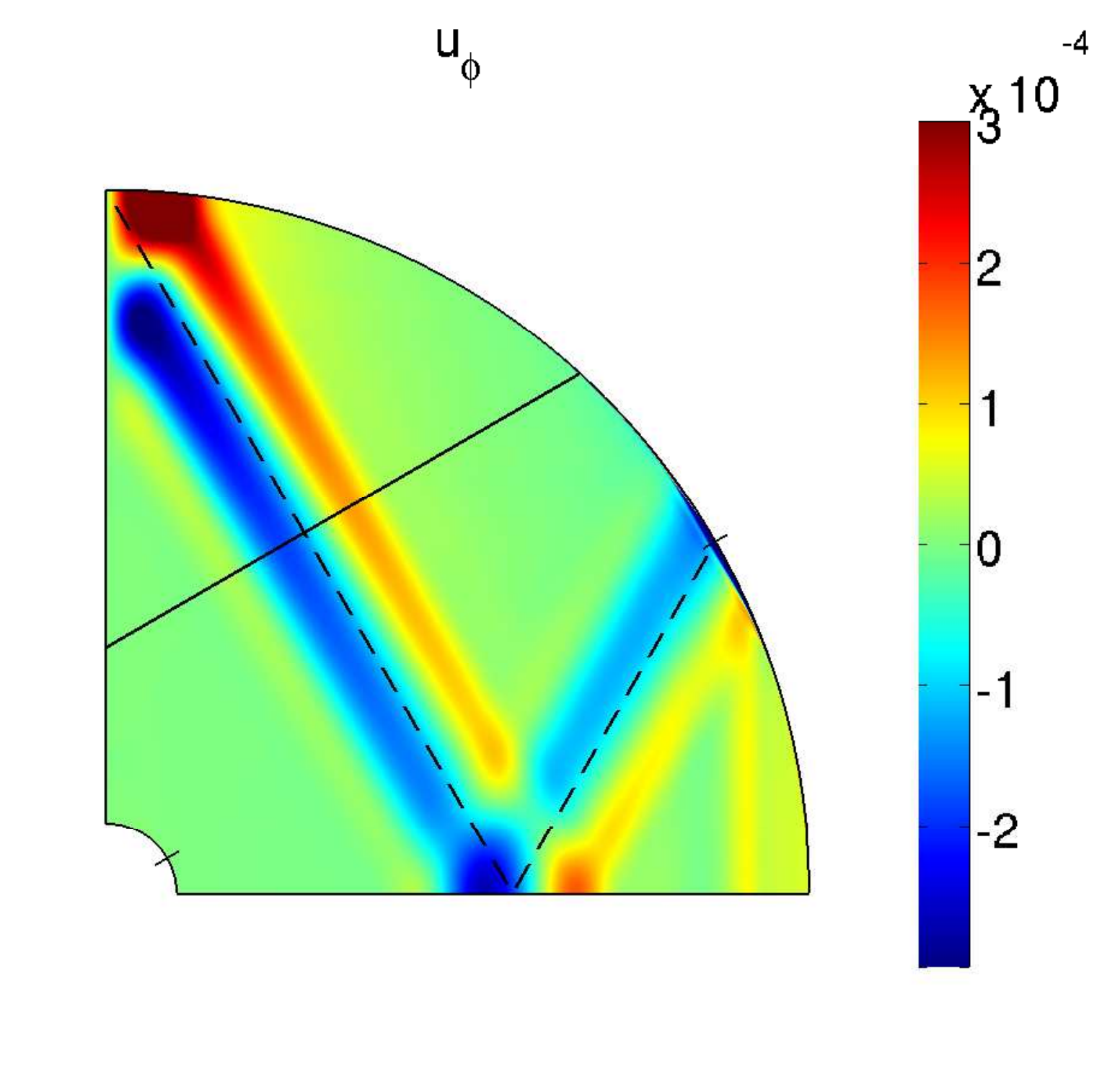}}
\hspace{0.2cm}
\subfloat[]{\includegraphics[scale=0.45]{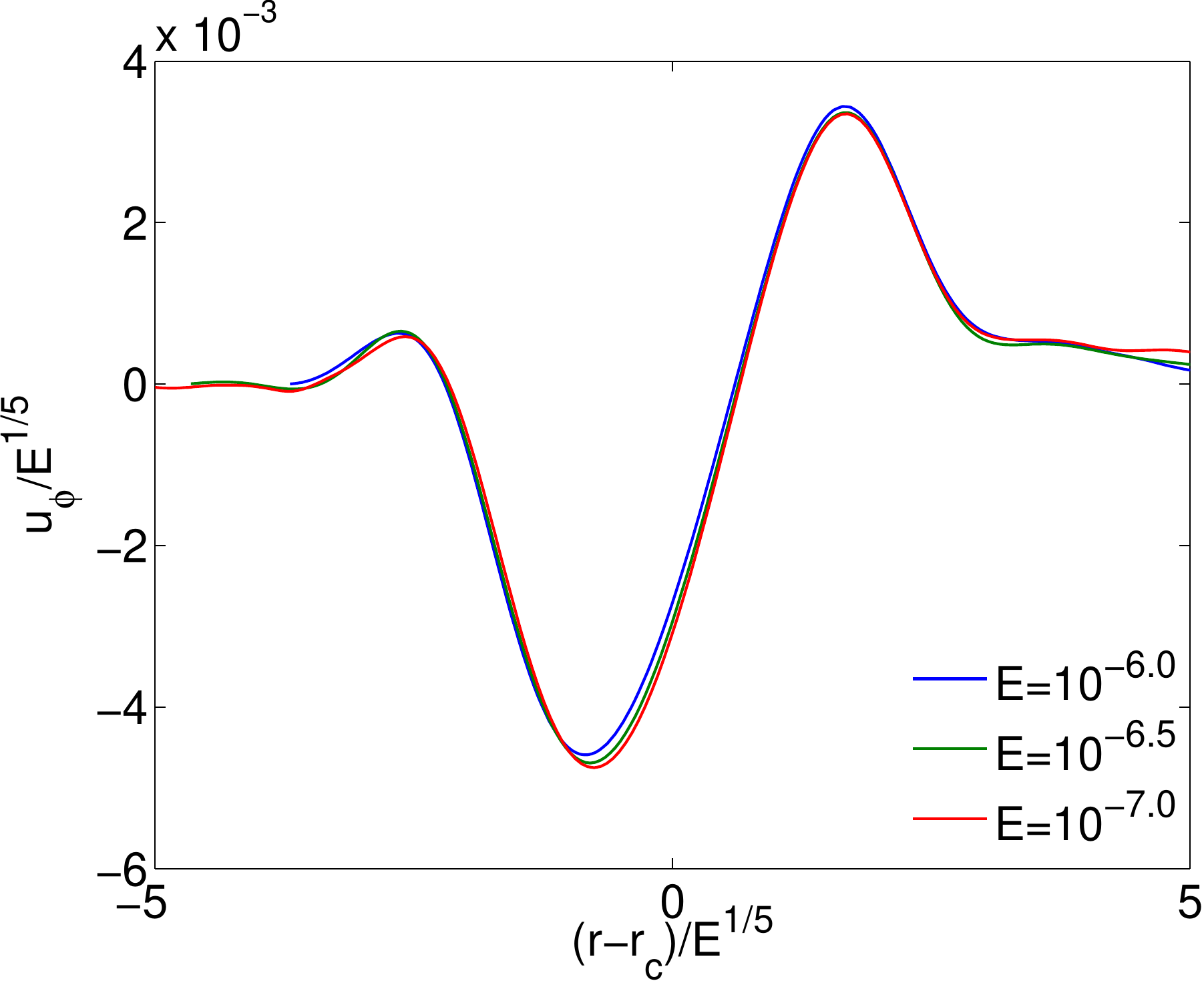}}
\caption{Azimuthal velocities for librating outer shell and uniformly rotating inner core. (a) Azimuthal velocities in a meridional plane when the outer shell is at the maximum retrograde position. Tick marks show the location of the critical latitude and dashed lines show the ray path of inertial wave exited from critical latitude. $E=10^{-7.0}, f=1.0, \varepsilon=0.01, \eta=0.1$. (b) Velocity profiles along a line (solid line in (a)) across the internal shear layer for different Ekman numbers. The coordinate origin is set at the center of the shear layer, the distance and velocity are both normalized by $E^{1/5}$.}
\label{outer_shear}
\end{center}
\end{figure}

Now we consider the case where the outer shell is librating at frequency $f=1.0$ with amplitude $\varepsilon=0.01$ while the inner core is uniformly rotating at $\Omega_o$. The radius ratio $\eta=0.1$. The inertial waves are excited by the breakdown of Ekman boundary layer at outer critical latitude this time. \autoref{outer_shear} (a) shows azimuthal velocities in a meridional plane when the outer shell is at the maximum retrograde position for Ekman number $10^{-7}$. Dashed lines are characteristic surfaces. In \autoref{outer_shear} (b) azimuthal velocity for different Ekman numbers are plotted along a line (Solid line in \autoref{outer_shear} (a)) perpendicular to the internal shear layer. Normalizing the distance and amplitude by $E^{1/5}$ we show that the scalings obtained in precession study by \citet{Noir2001} can also apply to libration. 
%In this case the path of the internal shear layers is also very simple, so we can test the scaling as previously. 

%Precession study \citep{Noir2001}  showed that both the width and amplitude of the internal shear layer spawned from outer critical latitude scale as $E^{1/5}$. We consider a line across the internal shear layer and see the variation as Ekman number. We set the coordinate origin at the centre of the shear layer and both the distance and velocity are normalized by $E^{1/5}$. \autoref{outer_shear} (b) shows that our scalings are consistent  with previous studies \citep{Roberts1963,Noir2001}.            

%\subsubsection{Internal shear layers spawned from both inner and outer critical latitude}
%It should be mentioned that the internal shear layers spawned from both inner and outer critical latitude simultaneously even though only the outer shell is librating. If the internal shear layer spawned from outer critical latitude touched the inner critical region, the inner Ekman boundary would also breakdown. This generate internal shear layers from inner critical latitude. We can see from \autoref{inner_outer} that the internal shear layers are generated from both inner and outer critical latitude when the inner shell is at rest and the outer shell is librating at  $f=1.4142$ with amplitude $\varepsilon=1.4142\times 10^{-2}$. The previous width scalings of the internal shear layers are still validate but the amplitude scalings are more complex due to indirect breakdown.    
\subsection{Zonal flow}
\begin{figure}
\begin{center}
\subfloat[]{\includegraphics[scale=0.35]{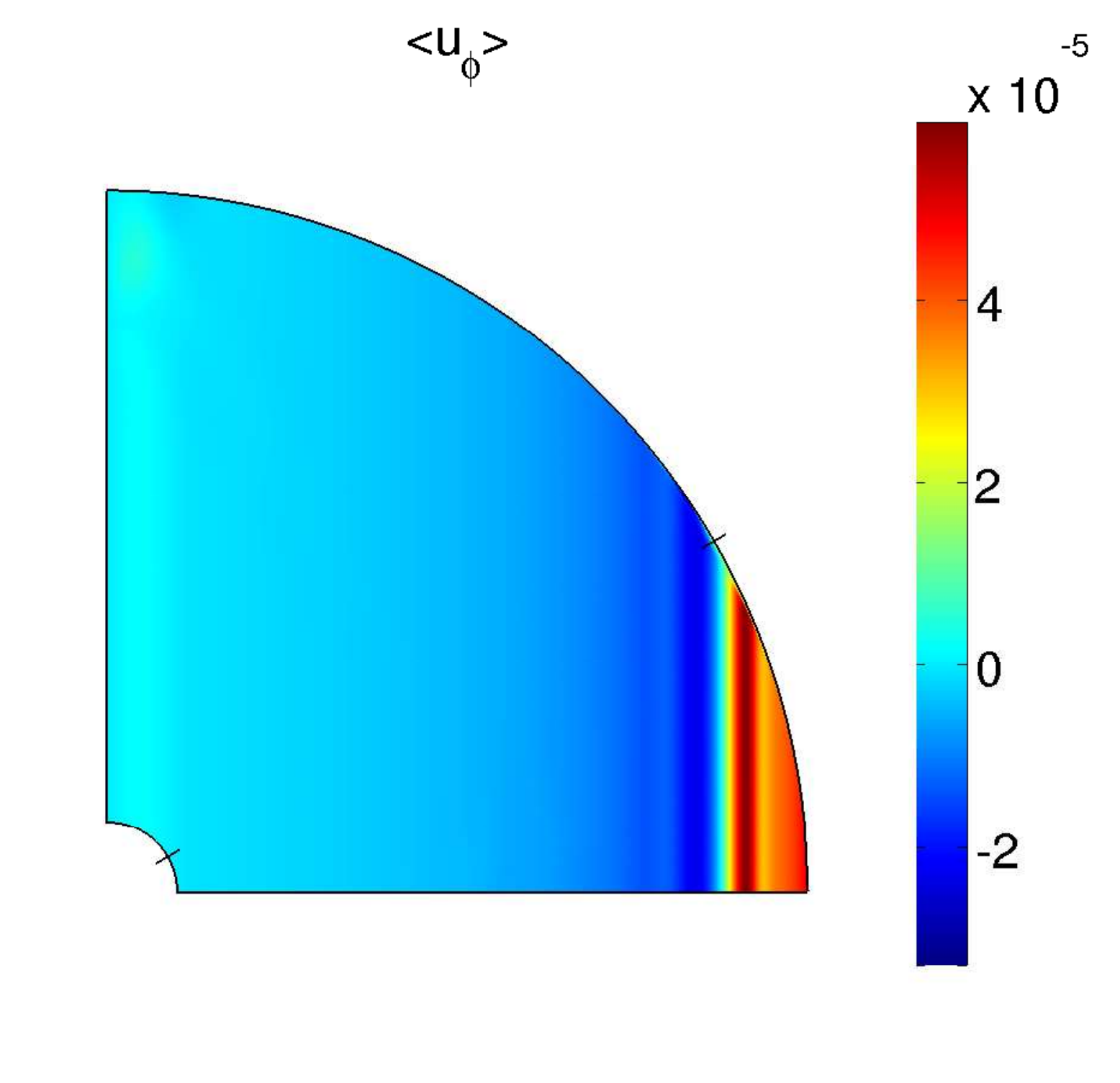}}
\subfloat[]{\includegraphics[scale=0.35]{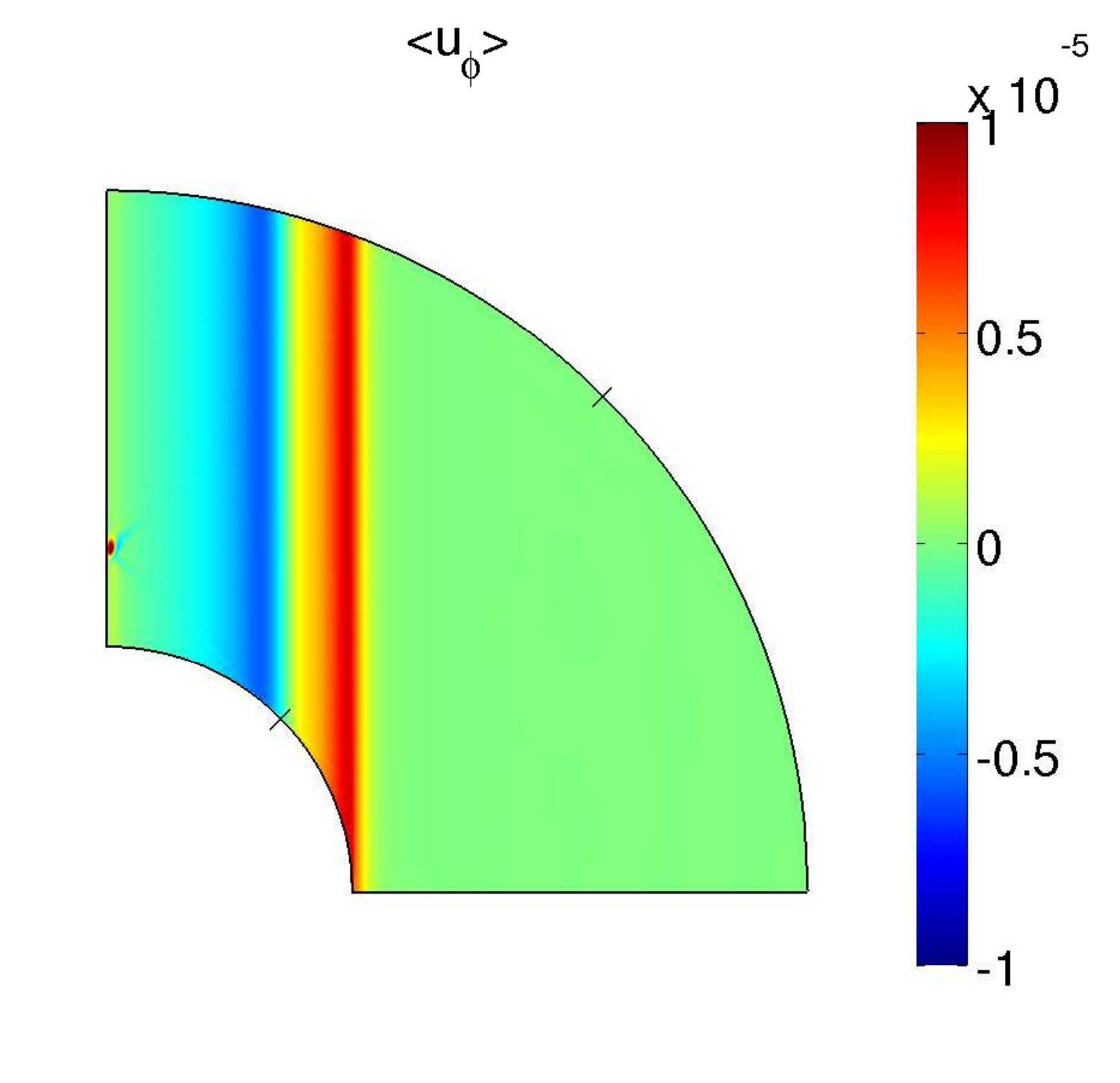}}
\subfloat[]{\includegraphics[scale=0.35]{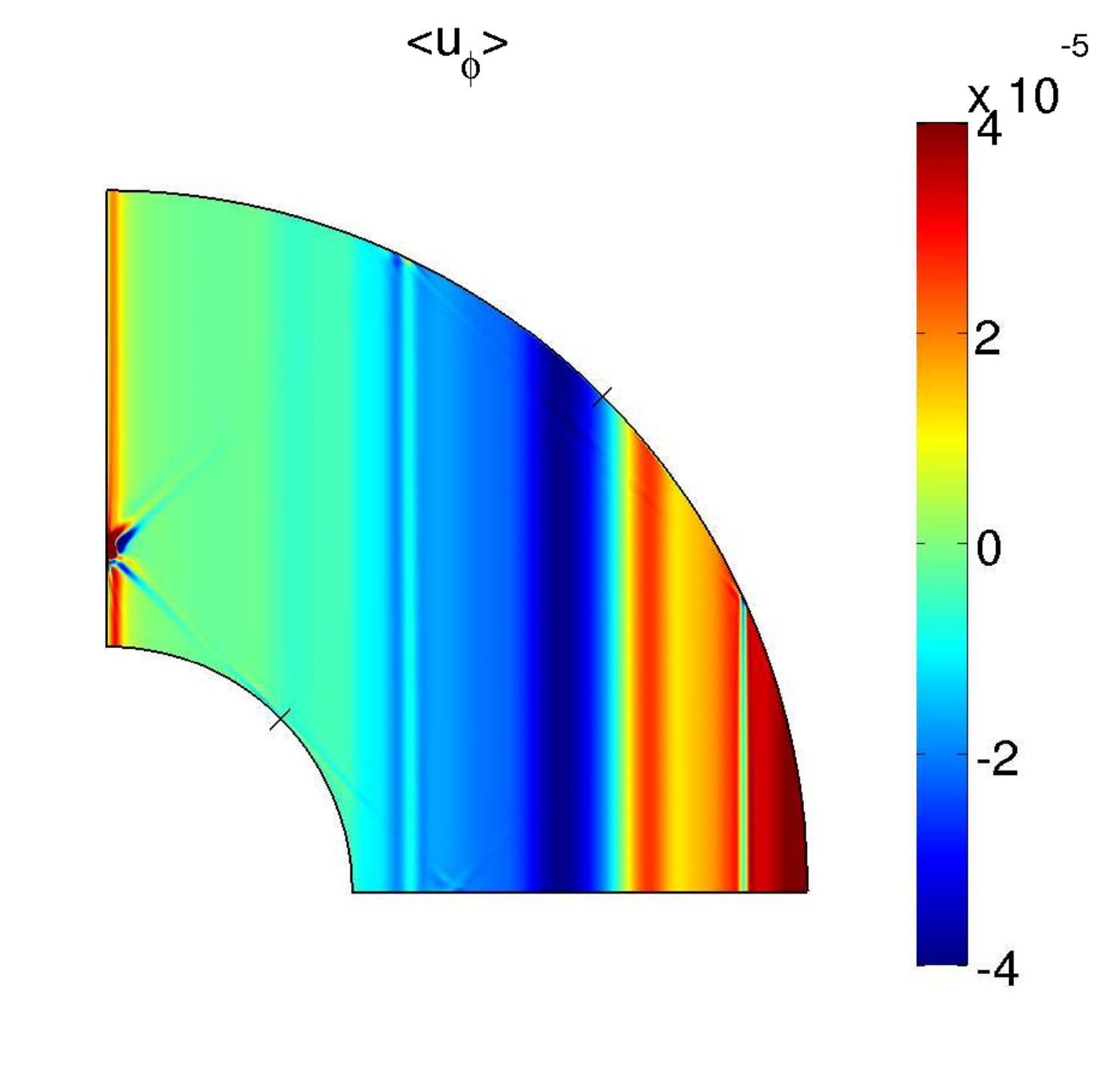}}
\caption{Time-averaged azimuthal velocities in a meridional plane. Tick marks show the location of the critical latitude. (a) The inner core is uniformly rotating at $\Omega_o$ and the outer shell is librating at frequency $f=1.0$, $\varepsilon=0.01$, $E=10^{-7.0}, \eta=0.1$;  (b) The outer shell is uniformly rotating at $\Omega_o$ and the inner core is librating at frequency $f=1.4142$, $\varepsilon=1.4142\times10^{-2}$, $E=10^{-7.0}, \eta=0.35$;  (c) The inner core is uniformly rotating at $\Omega_o$ and the outer shell is librating at frequency $f=1.4142$, $\varepsilon=1.4142\time10^{-2}$, $E=10^{-7.0}, \eta=0.35$.}
\label{zonal}
\end{center}
\end{figure}

\begin{figure}
\begin{center}
\subfloat[]{\includegraphics[scale=0.4]{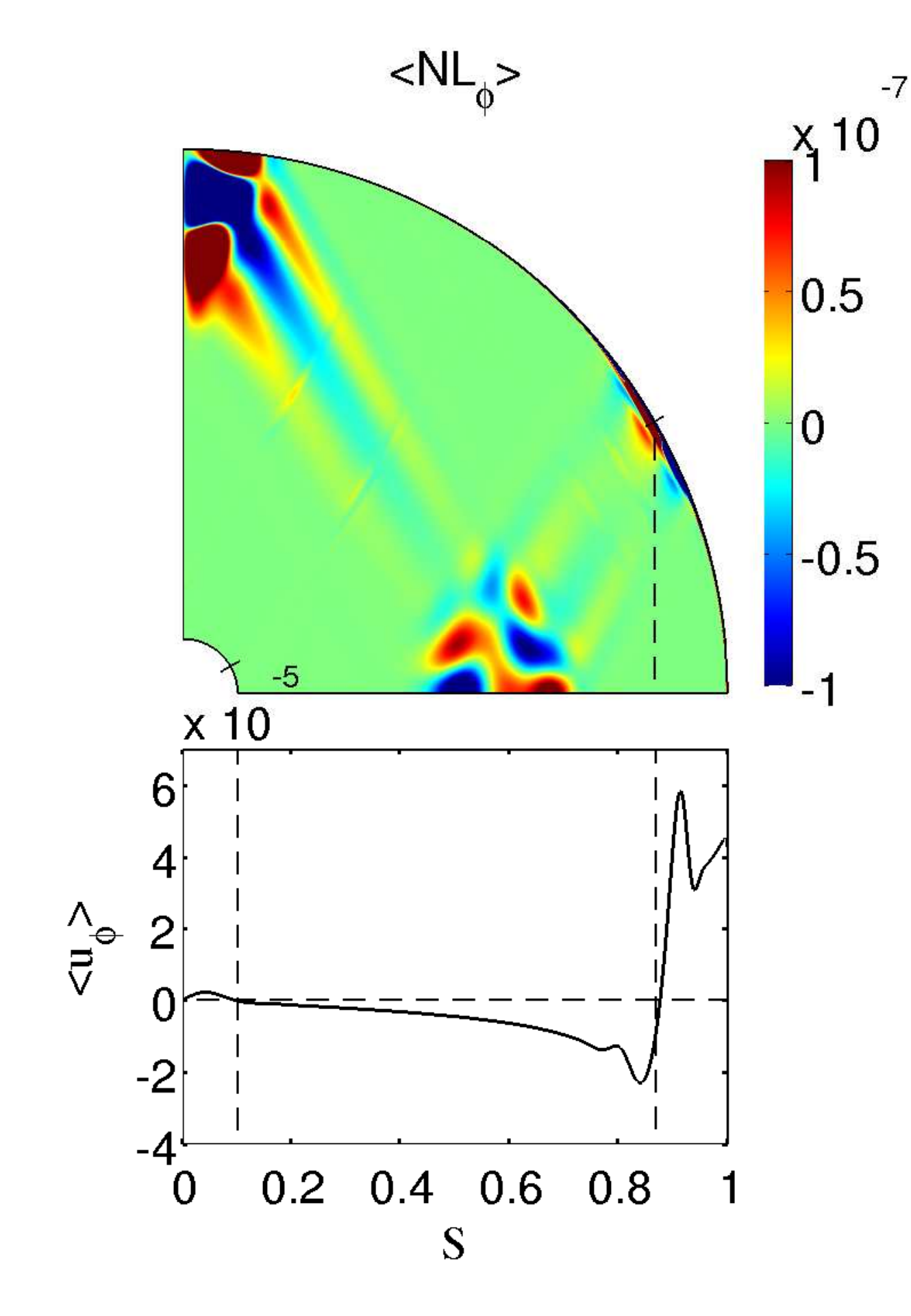}}
\subfloat[]{\includegraphics[scale=0.4]{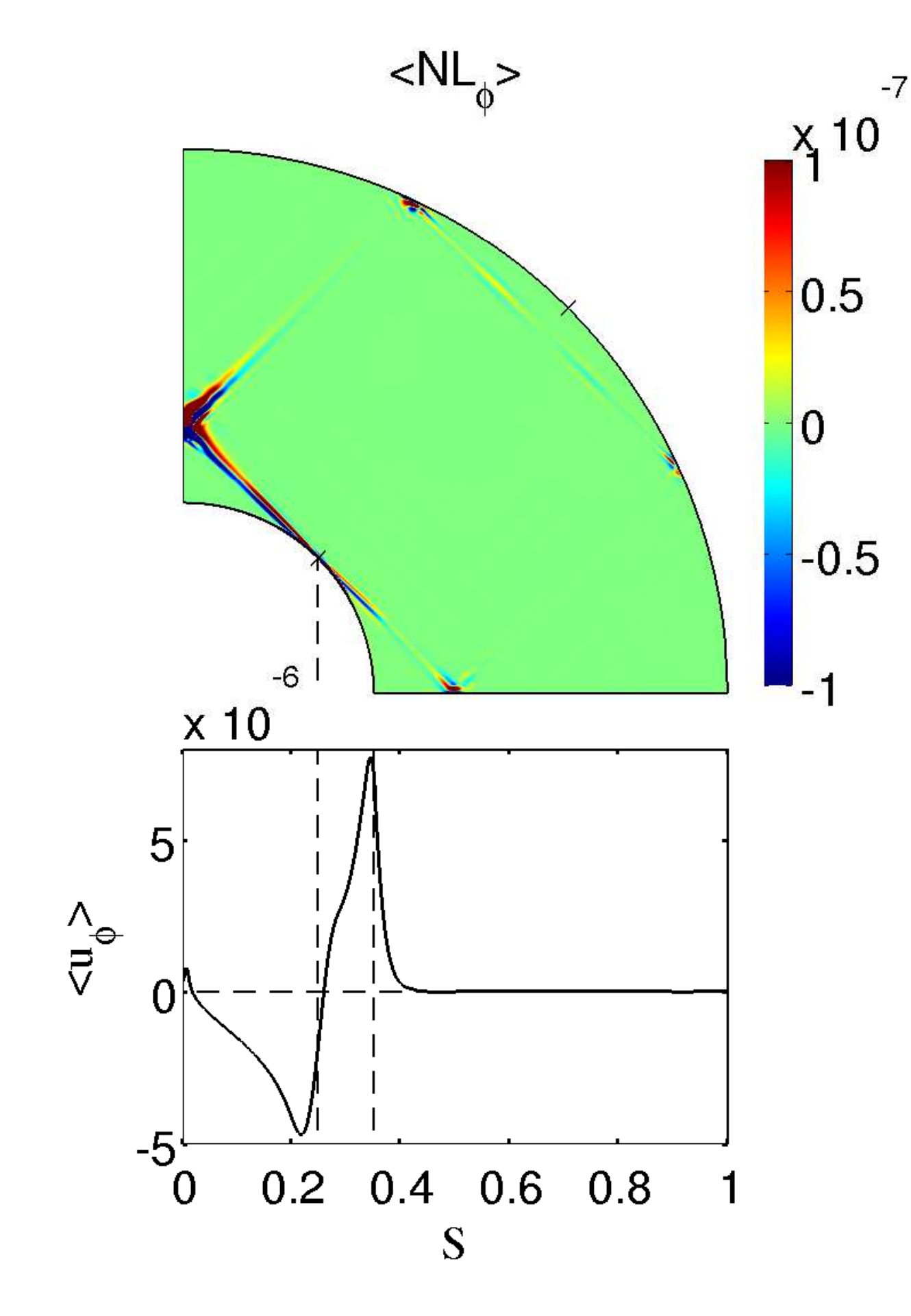}}
\subfloat[]{\includegraphics[scale=0.4]{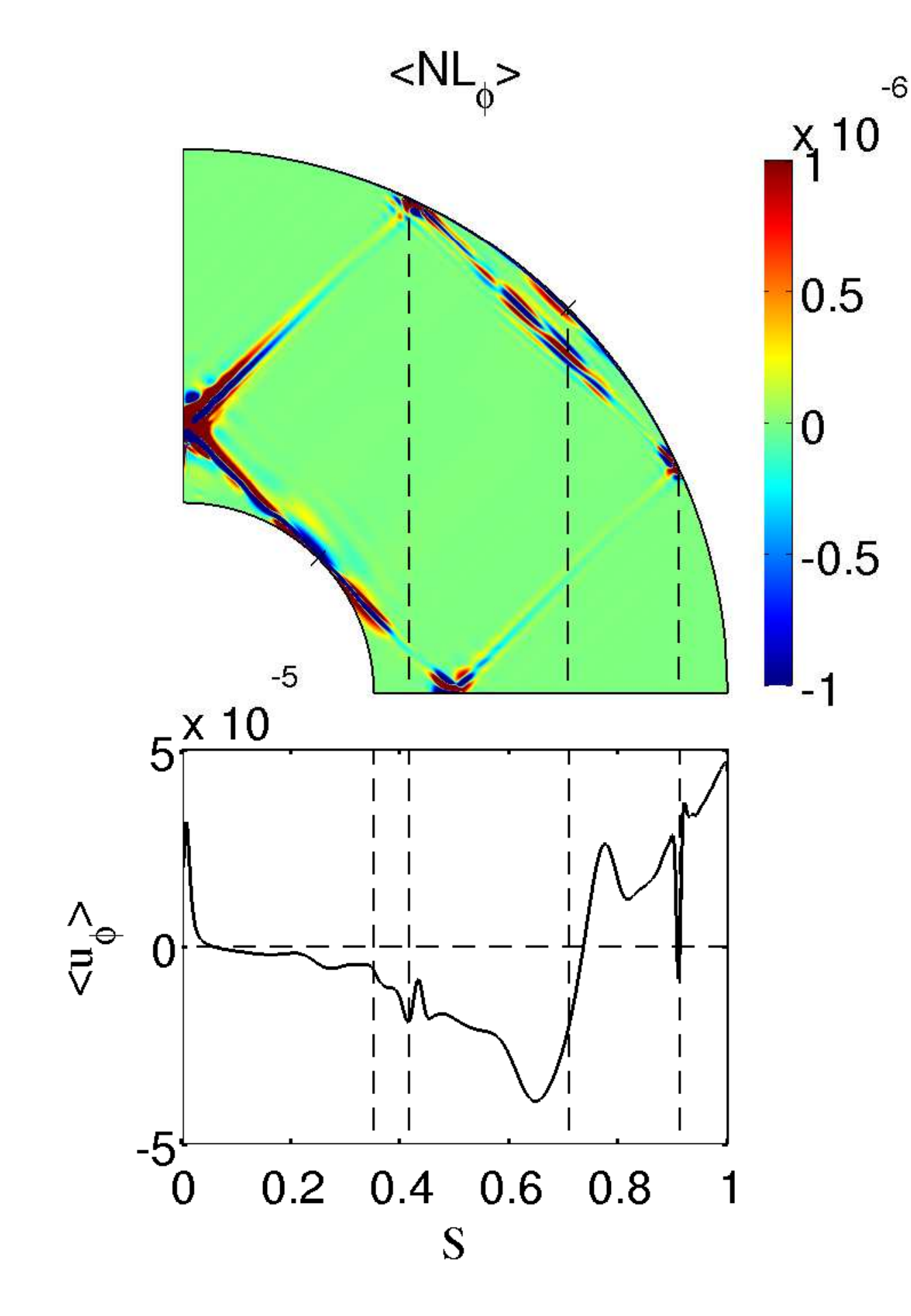}}
\caption{Time-averaged azimuthal non-linear force (colors) and geostrophic profiles (curves) corresponding to three cases in \autoref{zonal}. Dashed line marks inner radius or critical radius or cylindrical radius of reflection.}
\label{NLgeo}
\end{center}
\end{figure}

%Besides the time dependent flows, steady zonal flows are observed due to non-linearity in our simulations. 
In order to obtain the mean zonal flow, the azimuthal velocity is averaged in time over ten libration periods after reaching a stationary state. \autoref{zonal} shows three examples of zonal flows. (a) When the inner core is uniformly rotating at $\Omega_o$ and the outer shell is librating at frequency $f=1.0$ with amplitude $\varepsilon=0.01$, the zonal flows are dominated by a retrograde motion in the interior and a large prograde jet close to the outer equator. There is a significant geostrophic shear layer at outer critical cylindrical radius. (b) When the outer shell is uniformly rotating at $\Omega_o$ and the inner shell is librating at frequency $f=1.4142$ with amplitude $\varepsilon=1.4142\times10^{-2}$, there are no significant zonal flows outside the tangent cylinder. A geostrophic shear is observed at inner critical cylindrical radius. 
%This confirmed that the zonal flow is mainly driven by non-linear interaction in the Ekaman boundary layers. 
(c) When the inner core is uniformly rotating at $\Omega_o$  and the outer core is librating at $f=1.4142$ with amplitude $\varepsilon=1.4142\times10^{-2}$, a geostrophic shear is formed at outer critical cylindrical radius as previously. In addition, there are another two significant geostrophic shears. These two shear layers are probably generated by the reflection of internal shear layers on the boundary. In order to identify the underlying mechanism of the multiple geostrophic shears, \autoref{NLgeo} shows the time-averaged azimuthal non-linear force in a meridional plane (upper panels), and axially-averaged zonal velocities variation as cylindrical radius (low panels) for corresponding three cases in \autoref{zonal}. We can see from \autoref{NLgeo} (c) that besides the large geostrophic shear at critical radius, the other two strong geostrophic shears are exactly at cylindrical radius corresponding reflection of internal shear layers. The results suggest that the multiple geostrophic shear layers result from non-linear interaction in the boundary layers at the reflection points of the internal shear layers spawned from the inner critical latitude. Furthermore, in all three cases, although the non-linearities on the internal shear layers are dominated in the interior fluids, they don't drive significant zonal flows. This can be understood by deriving the torque due to non-linearities acting on a geostrophic cylinder. As seen from upper panels in \autoref{NLgeo}, although non-linear forces are significant in the bulk, they cancel out when integrated over the axial direction resulting in a negligible torque on a geostrophic cylinder. 
%  Seen from our numerical results, zonal flows are driven by non-linear interaction in the Ekman boundary layers. Both the breakdown of the Ekman boundary at critical latitude and internal shear layer reflection on boundary can generate geostrophic shear layers. 

\begin{figure}
\begin{center}
\includegraphics[scale=0.7]{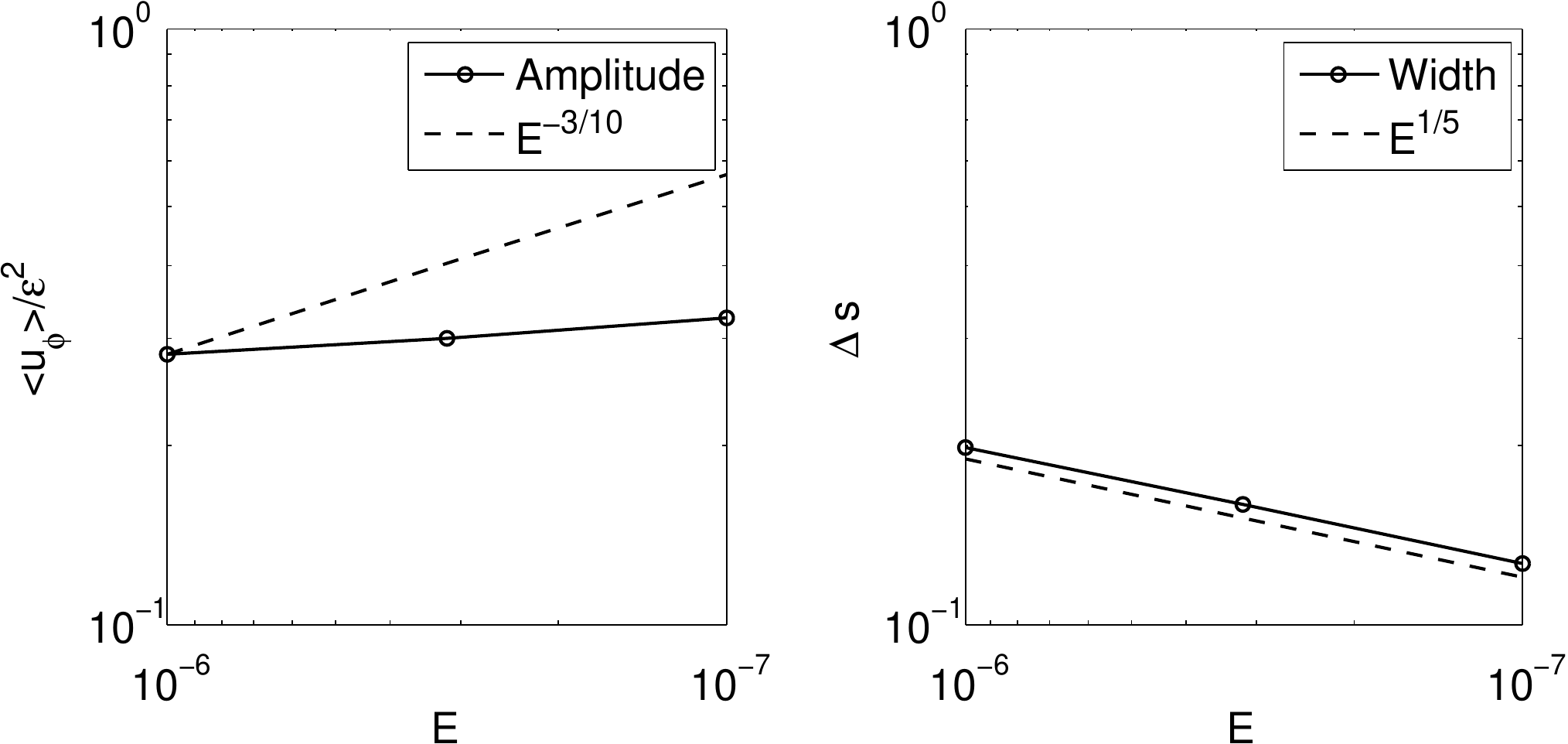}

(a) \hspace{7cm}(b)

\includegraphics[scale=0.7]{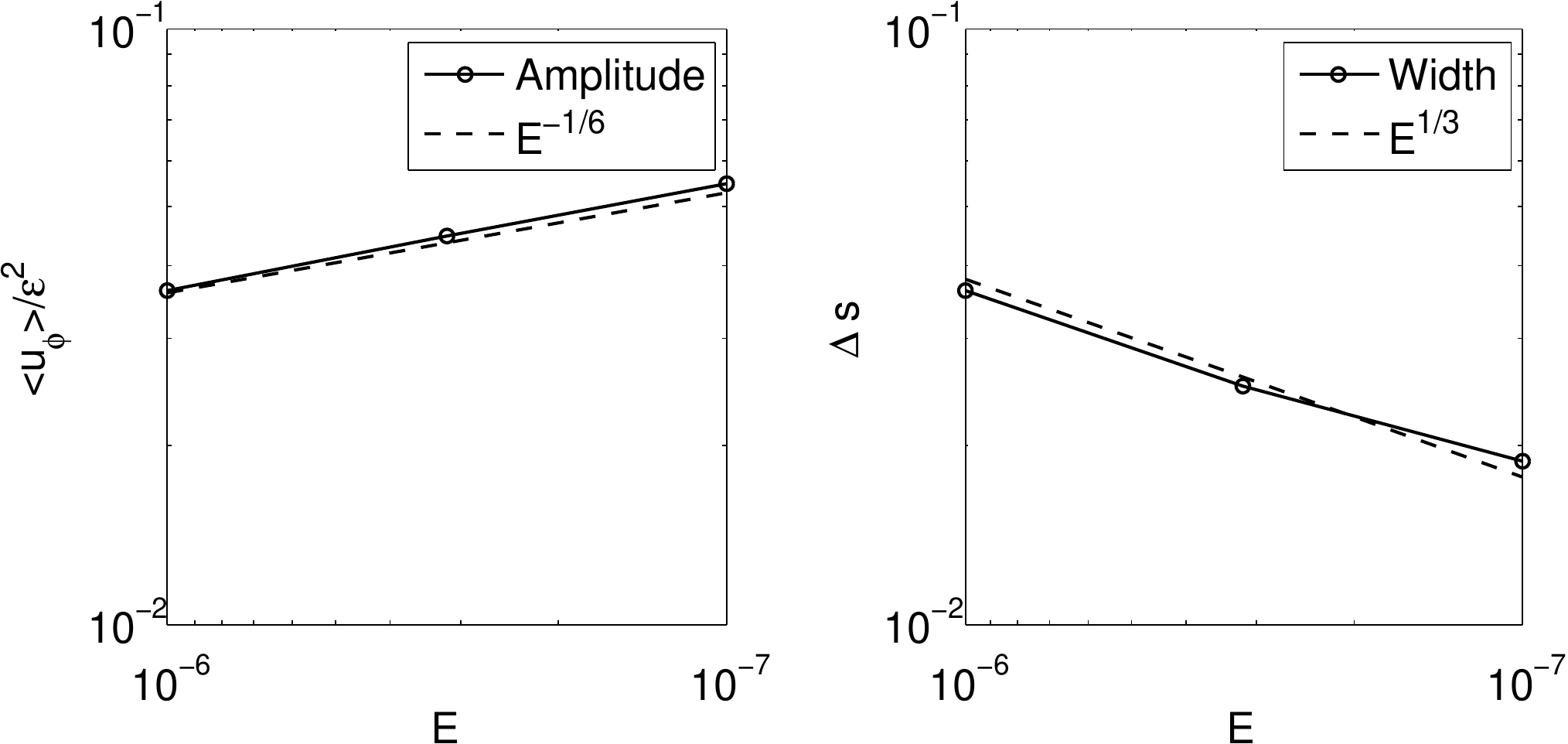}

(c) \hspace{7cm} (d)
\caption{(a) Amplitude and (b) width of geostrophic shear layer at outer critical latitude; (c) Amplitude and (d) width of geostrophic shear layer associated with reflection of internal shear layers.}
\label{geo_scale}
\end{center}
\end{figure}

In order to investigate the variation of geostrophic shear layer as Ekman number, we calculate the geostrophic flows in the case of \autoref{NLgeo} (c) but with Ekman numbers ranging from $10^{-6}$ to $10^{-7}$. We take the difference of maximum and minimum geostrophic velocity around the shear and distance between maximum and minimum as amplitude and width of geostrophic shear layer respectively. The amplitude and width of geostrophic shear layer are plotted as Ekman number in \autoref{geo_scale} . The width of the geostrophic shear layer at the critical cylindrical radius scales as $E^{1/5}$, which is the same as the width of breakdown region of the boundary layer at the critical latitude. However, the amplitude scaling is not consistent with the $E^{-3/10}$ obtained by precession study \citep{Noir2001}. It looks the amplitude of this geostrophic shear is beginning to saturate as decreasing the Ekman number. The amplitude and width of the geostrophic shear layer related to the reflection of internal shear layers scale as $E^{-1/6}$ and $E^{1/3}$ respectively. It is reasonable that the width scaling is same as the width of internal shear layers spawned from the inner critical latitude. The amplitude scaling of this kind of geostrophic shear will be discussed later from a viewpoint of torque balance acting on a geostrophic cylinder.    

%The amplitude of geostrophic shear layer at outer critical radius scales as $E^{-1/10}$ with width of $E^{1/5}$ (\autoref{geo_scale} (a) and (b)). The width scaling is the same as the width of the internal shear layers spawned form outer critical latitude. \autoref{geo_scale} (c) and (d) show the scalings of the geostrophic shear layer at reflection position close to the rotation axis. We can see that the amplitude and width of this geostrophic shear layer scales as  $E^{-1/6}$ and $E^{1/3}$ respectively. Although we only show the results of one geostrophic shear layer related the reflection of the internal shear layers, the other one has the same scalings. The width scaling of this kind of geostrophic shear layer is the same as the width of the internal shear layer spawned from inner critical latitude, as we showed previously. It's reasonable because this kind of geostrophic shear layer is generated by the reflection of internal shear layer spawned from inner critical latitude. We will discuss the amplitude scalings of geostrophic shear layers from a viewpoint of torque balance acting on a geostrophic cylinder later.                   

\section{Discussion}

%1. Torque balance between non-linear and viscous force acting on a geostrophic cylinder.

%2. Estimate the local Reynolds number in shear zones and apply to the planetary setting. 

%%%%%%%%%%%%%%%%%%%%%%%%%%%%%%%%%%%%%%%%%%%%%%%%%%%%%%%%%%%%%%%%%%%%%%%%%%%
\bibliographystyle{apalike} %{elsarticle-harv} %{abbrvnat} %{plainnat}
\bibliography{refrence_libration}

\newpage

%\begin{figure}
%\begin{center}
%\includegraphics[scale=0.5]{figure/w_path_14_o.pdf}
%\caption{Azimuthal velocity in meridian at the instant when the outer shell is at the maximum retrograde position, the inner shell is at rest. Tick marks show the location of the critical latitude and dashed lines show the ray path of inertial wave exited from critical latitude. $E=10^{-7.0}, f=1.4142, \varepsilon=1.4142\times10^{-2}$.}
%\label{inner_outer}
%\end{center}
%\end{figure}

\end{document}